\documentclass[iop,apj,numberedappendix,onecolappendix,revtex4-1]{emulateapj}  
\usepackage{amsmath}
\usepackage{graphicx}
\usepackage{epstopdf}
\usepackage[pdfa]{hyperref}
\usepackage{natbib}
\usepackage{color}
\usepackage{soul}
\hypersetup{colorlinks,
linkcolor=blue,
citecolor=blue,
filecolor=blue,
urlcolor=blue}

\usepackage{enumerate}

\newcommand{\dd}{\partial}
\newcommand{\del}{\mbox{\boldmath{$\nabla$}}}
\newcommand{\bcdot}{\mbox{\boldmath{$\cdot$}}}

\newcommand{\beq}{\begin{equation} }
\newcommand{\eeq}{\end{equation}}

\newcommand{\schwz}{ {\dd  \ln \frac{P}{\rho ^{\gamma}} \over \dd z}}
\newcommand{\schwR} { {\dd  \ln \frac{P}{\rho ^{\gamma}} \over \dd R} }

\begin{document}
\title{Fourier analysis of advection-dominated accretion flows}
\shorttitle{Fourier analysis of advection-dominated accretion flows}
\author{Asiyeh Habibi$^{1}$}
\author{Shahram Abbassi$^{1,2}$\altaffilmark{*}}
\author{Mohsen Shadmehri$^{3}$}
\affil{$^{1}$Department of Physics, School of Sciences, Ferdowsi University of Mashhad, Mashhad, 91775-1436, Iran;  \textcolor{blue}{abbassi@um.ac.ir}\\
$^{2}$School of Astronomy, Institute for Research in Fundamental Sciences (IPM), P.O. Box 19395-5531, Tehran, Iran\\
$^{3}$Department of Physics, Faculty of Sciences, Golestan University, Gorgan 49138-15739, Iran}
\begin{abstract}
We implement a new semi-analytical approach to investigate  radially self-similar solutions for the steady-state advection-dominated accretion flows (ADAFs). We employ the usual $\alpha$-prescription for the viscosity and  all the components of the energy-momentum tensor are considered. In this case, in the spherical coordinate, the problem reduces to a set of eighth-order, nonlinear differential equations with respect to the latitudinal angle $\theta$. Using the Fourier expansions for all the flow quantities, we convert the governing differential equations to a large set of nonlinear algebraic equations for the Fourier coefficients. Using the Newton-Raphson method we solve the algebraic equations and ADAF properties are investigated over a wide range of the model parameters. We also show that the implemented series are truly convergent. The main advantage of our numerical method is that it does not suffer from the usual technical restrictions which may arise for  solving ADAF  differential equations near to the polar axis. In order to check the reliability of the approach, we recover some widely studied solutions. We also introduce a new varying $\alpha$ viscosity model and new outflow and inflow solutions for ADAFs are presented using Fourier expansion series.

\end{abstract}
\keywords{Accretion - accretion discs -black hole physics - hydrodynamics.}
\section{Introduction}

Over recent decades, observations of the spectrum of the energetic astronomical objects provided a good motivation for many astronomers to investigate the phenomena responsible for enormous emitted energy of these objects, for example, see \cite{f89}, \cite{maki}, \cite{swa}. Theoretical models consistent with observational evidence have revealed that accretion process around massive stars or black holes is a plausible mechanism for producing such energetic radiations. The standard theory of the accretion disks, i.e. \cite{shs}, \cite{nt}, is a remarkably successful theory for understanding some of the observational features of the quasars, X-ray binaries and AGNs, for more details see \cite{lp}, \cite{wat}, \cite{frank}, \cite{kato}, \cite{af}. In the standard Shakura \& Sanyaev disk model (SSD), the energy released by the turbulent viscosity is radiated locally and the accretion flow becomes cool efficiently. The SSD model is appropriate for accreting systems where accretion rate is smaller than the Eddington accretion rate, $\dot{M}_{ED}$. At the same time, however, some observational evidences put some limitations on the validity of SSD model: e.g., the disk is too cool to emit hard X- and gamma-rays observed in black-hole objects. In mathematical point of view, this limitations are related to simplified assumptions in deriving the energy equation which they had been neglected the heat transport by advective motions in the accretion flow.  This mechanism is essential in the innermost disk structure, especially in black-hole accretion. A detailed analysis of the inner region of the disks is entirely indispensable, since most radiation of the disks originates from there. On the other hand it is even more important, since it gives rise to disk models that are distinct from the standard one.

An alternative model which is applicable to structures where the accretion rate is much smaller than $\dot{M}_{ED}$, is known as advection dominated accretion flow (ADAF), has been introduced by \cite{ny94}, \cite{ny95a}, \cite{ny95b} (hereafter NY94, NY95a, NY95b respectively), \cite{ich97}. In this model, the generated heat by turbulence is stored as entropy and can be transported with the flow towards the central part rather than being radiated away from the system immediately after generation. The ADAF model is of great interest because of its widespread applications in describing the low luminous AGNs \citep{lud}, the quiescent and hard states of black hole binaries \citep[and for the latest review see \citeauthor{yn14} \citeyear{yn14}]{yn2007, nmc08, ho08, yn11}. This model has been also applied to the supermassive black hole at our Galactic Center Sagittarius $A^{*}$ (Sgr $A^{*}$), see \cite{yqn}.

In order to improve our understanding of the physics of hot accretion flows, a lot of improvements have been proposed, including its multidimensional dynamics, disc-jet connection, radiation mechanisms and various astrophysical applications. For a review see \cite{yn14}. An interesting phenomenon associated with an accretion system, such as an ADAF is  launching of the winds or outflows. On the other hand, the structure of hot accretion flow is also remarkably affected by outflows, which carry huge amounts of mass, momentum and energy from the disk \citep{bu09, kawa, yuana, yuanb, bu13, sama}. There are some direct and indirect observational evidences to confirm the existence of outflows in different accreting systems, such as low-mass X-ray binaries \citep{fender04, mf06} and  AGNs \citep{tw01, pr09}.

One of the main properties of NY94 solutions is that the flow possesses a positive value for the Bernoulli parameter at the regions near to the pole which implies that the flow are susceptible to produced wind. Recent hydrodynamics (HD) and magnetohydrodynamics (MHD) simulations also confirmed emerging outflow in ADAFs \citep{spb, igab, ig2000, sp, mach, yuana, yuanb}. Extensive theoretical efforts have been made to understand outflow launching mechanism in ADAFs and its potential effects on its underlying accretion flows. For instance, see \cite{XC97}(hereafter XC97), \cite{bb99}, \cite{bb04}, \cite{xw05}, \cite{tm06}, \cite{ks13}, \cite{deg}, \cite{sama14}, \cite{sama}, \cite{sama17}, \cite{kh14}. However, there are some observations which imply that in some astronomical systems, e.g., cool-core clusters and several massive elliptical galaxies, outflow does not play an important role in the dynamics \citep{fabian11,allen}. Therefore, the study of hot accretion flow systems without outflow can be still important. There are several analytical and numerical solutions in the relevant literature studying ADAFs structure in one or two dimensions, for example see NY94; NY95a; \cite{ xw05, shad, zerat, habibi}. Among these solutions, NY95a is of particular importance. In this paper the authors presented numerical axisymmetric self-similar solutions for steady state ADAFs structure in spherical coordinates $(r, \theta, \phi)$. In their model self-similar scaling in the radial direction has been adopted while vertical structure have been found using two boundary conditions, relaxation method and numerically. They adopted standard model for viscosity in which the viscosity coefficient $\alpha$ is assumed to be constant. Furthermore they assume that latitudinal component of velocity is zero, i.e. $v_{\theta}=0$ and all components of viscous stress tensor are non zero. With these assumptions, they found two type of solutions satisfying different boundary conditions on the radial velocity at the rotational axis $\theta=0$. Following XC97 we have used Fourier expansion for vertical structure of the flow. Exploiting Fourier analysis give us a chance to simply imposed and control the boundary conditions. In this paper we will show that using Fourier analysis we will be able to find Nrayan \& Yi 1995 solutions precisely and furthermore we introduce some new solutions which have not been already reported in ADAFs literature. More specifically we present solutions with nonzero $v_{\theta}$ at the poles. This type of non-spherical inflow solutions does not exist in NY95 pioneer paper.

On the other hand, all of the semi-analytical solutions that have been used self-similar approach for structure of ADAFs, are not capable to explain the the total vertical structure of the flow. In fact, in a certain angle they encounter a singularity, which make them somehow inefficient to study the vertical structure \citep{sama}. To handle this problem, we introduced a fully analytical solution for the vertical structure of ADAFs \cite{habibi}. However that analytical solution was a no-wind solution using a restricted functionality for the viscosity coefficient $\alpha$. In the current paper focusing still on the full description of the vertical structure of ADAFs and generalizing our approach to include more general $\alpha$ descriptions, we use Fourier analysis and introduce a powerful tool for finding vertical structure of the flow uniquely by adding proper physics.

To do so, we keep all the components of the energy momentum tensor in our calculations. Also, we have adopted the $\alpha$-prescription for the viscosity. However we assume that in the spherical coordinate system $(r, \theta,\phi)$, $\alpha$ is a function of $\theta$. This can be achieved by assuming that the kinematic coefficient of viscosity depends on $r$ and does not vary with $\theta$. as we already mentioned, this study can be considered as a complementary paper to our previous study \cite{habibi}. Varying $\alpha$ models are motivated by three-dimensional simulation of hot accretion flows, for example see \cite{Penna}. In this context we introduce some new solutions which have not been already reported in the literature and discuss their physical relevance. 
 
The outline of the paper is as follows. In section \ref{equations}, the governing equations in the spherical polar coordinates for a steady state viscous flow with non-zero latitudinal velocity are presented. Furthermore, we introduce a varying $\alpha$ model in this section. Then assuming appropriate radial self-similar solutions for all the quantities, we find an eighth-order system of differential equations for latitudinal part of the functions. Then we introduce proper physical boundary conditions for this system of differential equations. In section \ref{saa}, we briefly discuss the Fourier expansion method and the numerical procedure for finding a unique solution for the large nonlinear set of algebraic equations. Then, in section \ref{vzero} we use this approach for an ADAF system with zero $v_{\theta}$ and find four type of solutions. In fact, it turns out that there are only two free parameters in the system: the viscosity parameter $\tilde{\alpha}$, and the thermodynamic parameter $\epsilon$. Without loss of generality we fix $\tilde{\alpha}$ and vary $\epsilon$ to classify the solutions. We discuss the physics of the solutions and compare them with the solutions available in the literature. 
In section \ref{ds} we check the dynamical stability of the solutions and find regions which are convectively unstable. Furthermore, we discuss if this instability can change the global aspects of the solutions and produce outflow in the system. In section \ref{vnzero} we apply this method to a ADAF system with non-zero $v_{\theta}$. We show that, in this case, this method fails to find unique and convergent solutions. Finally, we summarize the results in section \ref{dis}. As a test for the reliability of the approach, we apply it to well-known solutions presented in NY95 for ADAFs with zero $v_{\theta}$ and exactly recover all of that solutions. We show the results for this case in the Appendix B.

\section{General Formulation}\label{equations} 
In this section, we introduce the governing differential equations and discuss the appropriate boundary conditions. More specifically, we present the basic equations in subsection \ref{bequations}. In subsection \ref{ln} we introduce a new model for viscosity in which $\alpha$ is not a constant. The relevant radial self-similar forms are presented in the subsection \ref{selfs}. We discuss the boundary conditions in the subsection \ref{bcond}. Let us start with the basic equations: 
\subsection{Basic Equations}\label{bequations}
We use the spherical polar coordinates $ (r, \theta, \phi) $, where $ r $ is the radial distance, $ \theta $ and $ \phi $ are the polar and the azimuthal angles, respectively. A black hole with mass $M$ is at the origin and its gravitational potential is $ \psi(r)= -GM/r $. Furthermore, we assume that the system is axisymmetric and is in a steady state configuration. 

It is important to note that the fluid is assumed to be viscose and we adopt the $\alpha$ viscosity description for viscosity \citep{ig96, spb, igab, ig2000, fa05, shad}. Furthermore, we keep all the components of the energy-momentum tensor, $T_{\mu\nu}$, throughout this paper.

The governing equations are the continuity equation, momentum equation, and the energy equation. These equations can be found in the hydrodynamics textbooks \citep[e.g.,][]{mm}. For the sake of completeness, however, we rewrite them here. Therefore, the continuity equation becomes
\begin{equation}\label{continuity}
	\frac{1}{r^{2}}\frac{\partial}{\partial r} \Big( r^{2} \rho v_{r} \Big)  + \frac{1}{r \sin \theta} \frac{\partial}{\partial \theta} \Big(\sin \theta \rho v_{\theta} \Big) = 0,
\end{equation}
and the three components of the momentum equations are written as
\begin{equation}\label{motion_r}
	v_{r} \frac{\partial v_{r}}{\partial r} + \frac{v_{\theta}}{r} \Big( \frac{\partial v_{r}}{\partial \theta} - v_{\theta} \Big) - \frac{v_{\phi}^{2}}{r} = - \frac{GM}{r^{2}} - \frac{1}{\rho} \frac{\partial p}{\partial r}+\frac{2\Psi}{\rho}
	\end{equation}
	\begin{equation}
	v_{r} \frac{\partial v_{\theta}}{\partial r} + \frac{v_{\theta}}{r} \Big( \frac{\partial v_{\theta}}{\partial \theta} +  v_{r}  \Big) - \frac{v_{\phi}^{2}}{r}  \cot \theta = \frac{1}{\rho r} \frac{\partial p}{\partial \theta}+\frac{2\Phi}{\rho}
	\end{equation}
	\begin{equation}
	v_{r} \frac{\partial v_{\phi}}{\partial r} + \frac{v_{\theta}}{r} \frac{\partial v_{\phi}}{\partial \theta} + \frac{v_{\phi}}{r} \Big( v_{r} + v_{\theta} \cot \theta \Big) = \frac{2\Gamma}{\rho},
\end{equation}
and the energy conservation equation is given by
\begin{equation}\label{energy}
	\rho \Big( v_{r} \frac{\partial e}{\partial r} + \frac{v_{\theta}}{r} \frac{\partial e}{\partial \theta} \Big) - \frac{p}{\rho} \Big(  v_{r} \frac{\partial \rho}{\partial r} + \frac{v_{\theta}}{r} \frac{\partial \rho}{\partial \theta} \Big)  = \frac{2f \Lambda}{\rho},
\end{equation}
Equations (\ref{continuity})-(\ref{energy}) are the main equations which with an equation of state make a complete set for finding the unknown functions. In the above equations the parameters $\Psi$, $\Phi$, $\Gamma$ and $\Lambda$ are defined as
\begin{gather}
\Psi=\frac{1}{r^2\sin \theta}\left[\frac{\partial}{\partial r}(r^2\sin \theta T_{rr})+\frac{\partial}{\partial\theta}(r \sin \theta T_{r\theta}) \right]\\ \nonumber~~~~~~~~~~~~~~~~~-\frac{T_{\theta\theta}+T_{\phi\phi}}{r}
\end{gather}
\begin{gather}
\Phi=\frac{1}{r\sin \theta}\left[\frac{\partial}{\partial r}(r\sin \theta T_{r\theta})+\frac{\partial}{\partial\theta}(\sin \theta T_{\theta\theta}) \right]\\\nonumber~~~~~~~~~~~~~~~~~~-\frac{T_{\phi\phi}\cot \theta-2T_{r\theta}}{r}
\end{gather}
\begin{gather}
\Gamma=\frac{1}{r}\left[ \frac{\partial}{\partial r}(r T_{r\phi}+\frac{\partial T_{\phi\theta}}{\partial\theta})\right] +\frac{2}{r}(T_{r\phi}+T_{\phi\theta}\cot\theta)
\end{gather}
\begin{gather}
\Lambda=\frac{\partial v_{r}}{\partial r}T_{rr}+\frac{\partial v_{\theta}}{\partial r}T_{r\theta}+\frac{\partial v_{\phi}}{\partial r}T_{r\phi}+\frac{1}{r}\Big(\frac{\partial v_{r}}{\partial \theta}-v_{\theta}\Big)T_{r\theta}\\ \nonumber +\frac{1}{r}\Big(\frac{\partial v_{\theta}}{\partial \theta}+v_{r}\Big)T_{\theta\theta}+\frac{1}{r}\frac{\partial v_{\phi}}{\partial \theta}T_{\theta\phi}-\frac{v_{\phi}}{r}(T_{r\theta}+T_{\theta\phi}\cot\theta)\\\nonumber~~~~~~~~~+\frac{T_{\phi\phi}}{r}(v_r+v_{\theta}\cot\theta)
\end{gather}
Here, $\rho $ is the gas density, $ p $ is the gas pressure, and, $ v_{r} $, $ v_{\theta} $ and $ v_{\phi} $ are the radial, azimuthal and toroidal components of the gas velocity. We also introduce $f$ as a fraction of the advected energy. Furthermore, $ e $ denotes the specific internal energy of the fluid and can be written as $e = p/\rho \Big(\gamma - 1 \Big)$, where $ \gamma \equiv c_{p}/c_{v} $ is the ratio of specific heats. The independent components of the energy- mommentum tensor in the polar spherical coordinate are given by
\begin{equation}
T_{rr}=\mu\frac{\partial v_r}{\partial r}-\frac{\mu}{3}\left[ \frac{1}{r^2}\frac{\partial (r^2 v_r)}{\partial r}+\frac{\frac{\partial}{\partial \theta}(\sin\theta v_{\theta})}{r \sin\theta}\right],
\end{equation}
\begin{equation}
T_{\theta\theta}=\mu\frac{\partial v_{\theta}}{\partial \theta}+\frac{\mu v_r}{r}-\frac{\mu}{3}\left[ \frac{1}{r^2}\frac{\partial (r^2 v_r)}{\partial r}+\frac{\frac{\partial}{\partial \theta}(\sin\theta v_{\theta})}{r \sin\theta}\right],
\end{equation}
\begin{equation}
T_{\phi\phi}=\frac{\mu v_r}{r}+\frac{\mu v_{\theta}}{r}\cot\theta-\frac{\mu}{3}\left[ \frac{1}{r^2}\frac{\partial (r^2 v_r)}{\partial r}+\frac{\frac{\partial}{\partial \theta}(\sin\theta v_{\theta})}{r \sin\theta}\right],
\end{equation}
\begin{equation}
T_{r\theta}=\frac{\mu}{2}\left[\frac{1}{r}\frac{\partial v_r}{\partial \theta}+r\frac{\partial }{\partial r}\Big(\frac{v_{\theta}}{r}\Big) \right],
\end{equation}
\begin{equation}
T_{r\phi}=\frac{r\mu}{2}\frac{\partial}{\partial r} \Big( \frac{v_{\phi}}{r}\Big),
\end{equation}
\begin{equation}
T_{\theta\phi}=\frac{\mu\sin\theta}{2r}\frac{\partial}{\partial \theta}\Big(\frac{v_{\phi}}{\sin\theta}\Big)
\end{equation}
Hereafter we define the isothermal sound speed as $c_s^2=p/\rho$ and the net mass accretion rate is given by
\begin{equation}\label{nac}
\dot{M}=-\int 2\pi r^2 \sin \theta \rho(r,\theta) v_r(r,\theta)d\theta
\end{equation}
\subsection{A latitudinally varying $\alpha$ viscosity model}\label{ln}
Our goal is to solve equations (\ref{continuity})-(\ref{energy}) using our numerical method to illustrate its ability and possible weaknesses. In doing so, we consider two cases with different viscosity prescriptions:
(i) First, we investigate an accretion flow with the commonly used $\alpha$-prescription where the kinematic coefficient of viscosity is $\nu=\alpha c_s^2(r,\theta)/\Omega_K(r)=\alpha r^2 \Omega_K(r)c_s^2(\theta)$. Where we have used the similarity solutions introduced in the next subsection, see equation \eqref{nee}. Here, the Keplerian angular velocity is $\Omega_K(r)=\sqrt{\frac{MG}{r^3}}$. Note that viscosity coefficient $\alpha$ is given model parameter and the viscosity is dependent on both the radial distance and the latitudinal angle. (ii) As a second illustrative configuration, we assume that viscosity is independent of the latitudinal angle. In this particular case, radially self-similar solutions are found if the kinematic viscosity coefficient is written as
\begin{equation}
\nu(r)=\tilde{\alpha}\,r^2\,{\Omega_{K}(r)}
\label{newnu}
\end{equation}
where $\tilde{\alpha}$ is a model parameter. In both cases (i) and (ii) the bulk viscosity, i.e., $\mu(r,\theta)=\rho(r,\theta)\nu$ depends upon $r$ and $\theta$. In the case (i), bulk viscosity $\mu$ is proportional to pressure, i.e., $\mu(r,\theta)=\alpha p(r,\theta)/\Omega_K(r)$, whereas in the case (ii) bulk viscosity $\mu$ is dependent on  the density, i.e., $\mu(r,\theta)=\tilde{\alpha}\rho(r,\theta)r^2\,\Omega_K(r)$.

Functional form of the introduced viscosity in case (ii) is actually a variant of the widely used $\alpha$-prescription if the viscosity coefficient $\alpha$ is assumed to be dependent on the latitudinal angle as follows
\begin{equation}
\alpha(\theta)=\frac{\tilde{\alpha}}{c_s^2(\theta)}.
\label{newal}
\end{equation}
Although this prescription seems to be restrictive, there are hot accretion flow simulations that suggest viscosity coefficient $\alpha$ is not a constant and it depends on the latitudinal angle \citep[e.g.,][]{Penna}. This dependence, however, is needed to be explored using further hot accretion flow simulations. But it motivated some authors to explore properties of ADAFs with viscosity parameter as a given function of the latitudinal angle. \cite{habibi}, for instance, found fully analytical no-wind solutions for the ADAFs with viscosity parameter in proportion to $\sin\theta$. In their analysis, however, the radial-azimuthal component of the stress tensor, i.e., $T_{r\phi}$ is assumed to be dominant. This assumption is relaxed in our analysis by fully implementing all the stress components. Our solutions with the introduced viscosity prescription in case (ii) are  direct generalizations to the analysis of  \cite{habibi}.

In the next subsections, we present similarity solutions for  ADAFs with viscosity as introduced in cases (i) and (ii).

\subsection{Self-similar Solutions}\label{selfs}
By assuming radial self-similar solutions, one may convert the two-dimensional differential equations (\ref{continuity})-(\ref{energy}) to a set of one-dimensional differential equations. To do so, we follow NY95 by using the following radial dependencies for the quantities
\begin{gather}
	\rho(r,\theta) = r^{-n} \rho(\theta) ,\\\label{nee}
	c_s(r, \theta) = \sqrt{\frac{MG}{r}}c_s(\theta),\\
	v_{r}(r,\theta) = \sqrt{\frac{GM}{r}} v_{r}(\theta),\\
	v_{\theta}(r, \theta) = \sqrt{\frac{MG}{r}} v_{\theta}(\theta),\\
	v_{\phi}(r, \theta) =   \sqrt{\frac{GM}{r}} \sin\theta \, \Omega(\theta),
\end{gather}
Furthermore we define the function $q(\theta)$ to include both models A and B. To do so we express $\mu$ as $\mu(r,\theta)=\rho(r,\theta)r^2 \Omega_K q(\theta)$. The function $q(\theta)$ is equal to $\alpha c_s^2(\theta)$ for model A, and is equal to $\tilde{\alpha}$ in model B.

Upon substituting the above solutions into equations (\ref{continuity})-(\ref{energy}),  the following ordinary differential equations are obtained:
\begin{equation}\label{1}
(3-2 n) \rho  v_r+2 v_{\theta } \Big(\rho '+\rho  \text{cot$\theta $}\Big)+2 \rho  v_{\theta }'=0
\end{equation}
\begin{equation}\label{2}
\begin{split}
\rho & \Big(6 (n+1) c_s^2+ q \Big(12 (n-2) v_r+  (4 n-17) v_{\theta }'\\&+(4 n-17) \cot \theta  v_{\theta }+6 \cot \theta  v_r'+6 v''(\theta )\Big)\\&+3 \Big(2 \Omega (\theta )^2 \sin ^2(\theta )-2 v_r' \Big(v_{\theta }- q'\Big)-3   q' v_{\theta }\\&+v_r^2+2 v_{\theta }^2-2\Big)\Big)+3  q \rho ' \Big(2 v_r'-3 v_{\theta }\Big)=0
\end{split}
\end{equation}
\begin{equation}\label{3}
\begin{split}
\rho & \Big(6 \left(c_s^2\right)'-6 \Omega (\theta )^2 \sin \theta  \cos \theta +3 v_r \Big(v_{\theta }-2  q'\Big)\\&+2 v_{\theta }' \Big(3 v_{\theta }-4   q'\Big)+4   \cot \theta  q' v_{\theta }\Big)+6 c_s^2 \rho '\\&-  q \Big(\rho  \Big(-6 (n-3) v_r'+v_{\theta } \Big(-8 \csc ^2\theta +9 n-6\Big)\\& +8 \Big(\cot \theta v_{\theta }'+v_{\theta}''\Big)\Big)+2 \rho ' \Big(3 v_r+4 v_{\theta }'-2 \cot \theta  v_{\theta }\Big)\Big)=0
\end{split}
\end{equation}
\begin{equation}\label{4}
\begin{split}
  q &\Big(-2 \sin \theta  \rho ' \Omega '-\rho  \Big(6 \cos \theta  \Omega ' +\sin \theta  \Big(2 \Omega ''\\&+3 (n-2) \Omega \Big)\Big)\Big)+\rho  \Big(2 \sin \theta  \Omega ' \Big(v_{\theta }-  q'\Big)\\&+\Omega \sin \theta  v_r+4 \Omega \cos \theta  v_{\theta }\Big)=0
\end{split}
\end{equation}
\begin{equation}\label{5}
\begin{split}
36 \epsilon & c_s^2 \Big(\rho  ((\gamma -1) n-1) v_r-(\gamma -1) \rho ' v_{\theta }\Big)\\&+\rho  \Big(36 \epsilon  v_{\theta } \Big(c_s^2\Big)'+\frac{1}{2}  (3 \gamma -5) q \csc ^2\theta  \\& \Big(54 \Omega^2 \sin ^4\theta +8 \sin ^2\theta \Big(3 \sin ^2\theta  \Omega '^2+3 \Big(v_r'\Big){}^2\\&+4 \Big(v_{\theta }'\Big){}^2\Big)-8 \sin \theta  v_{\theta } \Big(9 \sin \theta  v_r'+4 \cos \theta  v_{\theta }'\Big)\\&+48 \sin \theta  v_r \Big(\sin \theta  v_{\theta }'+\cos \theta  v_{\theta }\Big)\\&+72 \sin ^2\theta  v_r^2+(43-11 \cos 2 \theta ) v_{\theta }^2\Big)\Big)=0
\end{split}
\end{equation}
where prime stands for the derivative with respect to $\theta$ and the new parameter $\epsilon$ is defined as $\epsilon=(5/3-\gamma)/(\gamma -1)f$. Equations (\ref{1})-(\ref{5}) are the main non-linear differential equations which we shall solve in the subsequent sections for five unknowns $\rho$, $v_r$, $v_{\theta}$, $\Omega$ and $c_s$. It is necessary to emphasis that in the case (i) where $q(\theta)=c_s^2(\theta)$, our equations are slightly different from equations B1-B5 presented in \cite{xw05}, and, implemented equations in XC97. We found some minor mistakes in \cite{xw05}. If we set $v_{\theta}=0$ in \cite{xw05} equations, the NY95 equations are not recovered.

\subsection{Boundary Conditions}\label{bcond}

We now need a set of proper boundary conditions for solving the equations. Equations (\ref{1})-(\ref{5}) constitute an eighth-order system of ordinary differential equations. We, therefore, need eight boundary conditions to solve them numerically.  Definition of the net mass accretion rate provides the first boundary condition. In doing so, the integral (\ref{nac}) sets the normalization of $\rho(\theta)$, i.e., one should fix the magnitude of $\dot{m}=\dot{M}/(2\pi\sqrt{ MG})$. More specifically, in the case of $n=3/2$ this quantity does not depend on $r$ and provides a suitable boundary condition. We will use this boundary condition for all the solutions presented in this paper for $n=3/2$. Instead of this condition, one may use $\rho(\pi/2)=1$ because we can simply scale $\rho(\theta)$ by the accretion rate. We will use this condition for the cases where $n\neq 3/2$. One should note that for these solutions $\dot{m}$ can not be non-zero. In other words, the mass conservation implies that the net mass accretion rate is not dependent on $r$. On the other hand, the self-similar solutions yield $\dot{m}=I r^{3/2-n}$, where $I$ is an integral given by
\begin{equation}
I=-\int_0^{\pi}\rho(\theta)v_r(\theta)\sin \theta d\theta
\end{equation}



The remaining seven boundary conditions are distributed between the equatorial plane and the rotation axis $\theta=0$. At the equatorial plane, the boundary conditions can be written as 
\begin{equation}\label{bc1}
\theta=\frac{\pi}{2}:~~~\frac{d\rho}{d\theta}=\frac{d v_r}{d\theta}=\frac{d \Omega}{d\theta}=\frac{d c_s}{d\theta}=v_{\theta}=0
\end{equation}
The other two conditions can be obtained by fixing the magnitude of $v_{\theta}(\pi/2)$ and $c_{s}(\pi/2)$ as  has been done in Xue \& Wang (2005). In this case, we have an initial value problem. However by fixing these conditions at $\theta=0$ the system will be a boundary value problem. On the other hand, at $\theta=0$, we expect that the solutions be well behaved and nonsingular. In this case, we have
 \begin{equation}\label{bc2}
\theta=0:~~~\frac{d\rho}{d\theta}=\frac{d v_r}{d\theta}=\frac{d \Omega}{d\theta}=\frac{d c_s}{d\theta}=0
\end{equation}
By imposing these conditions into equations (\ref{1})-(\ref{5}) at $\theta=0$, one can easily verify that 
\begin{equation}
v_r=0~~~~~~~\text{or}~~~~~~v_r=-\frac{\epsilon\, c_s^2}{2 q}
\label{newbc}
\end{equation}
Obviously the number of conditions are larger than eight. However one should note that all of them are not independent. Technically, we chose a convenient eight component subset of the conditions in order to solve the equations.
\section{Semi-analytic approach: Fourier expansion}\label{saa}
In this section, we follow the method introduced in XC97. However, we use the correct set of equations and also use a different set of boundary conditions. We also find new solutions when the latitudinal angle dependence of the viscosity parameter $\alpha$ is permitted as in case (ii). The idea is that we can always construct proper Fourier series for the physical quantities which satisfy the boundary conditions (\ref{bc1}) and (\ref{bc2}) by default. More specifically in the simplest case we can express the quantities as the following series
\begin{equation}
\begin{split}
&\rho(\theta)=\sum_{i=0}^{N}a_i \cos 2 i \theta,~~~~~~~~c_s^2(\theta)=\sum_{i=0}^{N}w_i \cos 2 i \theta\\&
v_r(\theta)=\sum_{i=0}^{N}b_i \cos 2 i \theta,~~~~~~~\Omega(\theta)=\sum_{i=0}^{N}d_i \cos 2 i \theta\\&
v_{\theta}(\theta)=\sin 2\theta\,\sum_{i=0}^{N}h_i \cos 2 i \theta
\end{split}
\label{fs}
\end{equation}
Substituting these series into equations (\ref{1})-(\ref{5}) and using an appropriate subset of boundary conditions, we obtain a $5(N+1)$ nonlinear algebraic equations for $5(N+1)$ coefficients. In other words, the problem reduces to solve algebraic equations than the differential equations. Using the Newton-Raphson method, we solve these equations. In this method one needs appropriate primary guess for the solutions, We find them by using random number generators. In order to see how this procedure works, we have written the details for a toy model in which $\Omega=0$ in the Appendix A. In fact our equations are too long and we can not write the main calculations in the paper.

The practical power of this approach is that the boundary conditions have been already included in the series and consequently by finding the coefficients one may analytically analyze the quantities in the whole interval $[0,\pi]$. In other words, these solutions cover the whole space and allow to investigate the dynamics in a complete manner. Therefore this approach may help to better understanding of the ADAF structure.

However, the technical limitation is that we have to truncate the expansion in a specific $N$. In fact when $N$ is larger than $9$ the parameter space of solutions get extremely large and in practice, it is not possible to find a unique solution. We start with $N=3$ and increase $N$ by checking the convergence of the solutions. More specifically when the first and dominant coefficients in the series remain approximately constant by increasing $N$ we decide that the numerical procedure is convergent. More specifically in each step, we measure the fractional difference between the coefficients in $N$ and $N+1$ cases. Albeit it should be noted that increasing $N$ does not necessarily yield to better solutions. In fact for large N, $N>10$, the numeric errors dominate the calculation and the solution diverges and takes a highly oscillatory and unnatural form. In some cases, we use $N=9$ for the number of coefficients. Fortunately, in all cases we find an acceptable convergence for $5\leq N\leq 9$.
\section{Advection dominated accretion flows with $v_{\theta}(\theta)=0$}\label{vzero}
Our study is started with a simplified configuration with $v_{\theta}(\theta)=0$. Existence of similarity solutions imply that we have $n=3/2$ in both cases (i) and (ii).
\subsection{Model A: solutions with $\alpha$-prescription}
Accretion flow with a viscosity prescription introduced in case (i) is actually equivalent to the model studied in NY95. If we set $q(\theta)=\alpha c_s^2$ in the equations (\ref{1})-(\ref{5}), these equations reduce to equations (2.16)-(2.19) in NY95. To illustrate that the Fourier analysis method is an efficient  tool for instigating an ADAF structure, we first retrieve NY95 results in the Appendix B along with verification that the implemented series are truly convergent. Although we use a different numerical method, our obtained solutions are consistent with NY95 solutions with reasonable accuracy. This successful test problem is a good motivation to implement the Fourier analysis method for exploring ADAF structure with varying viscosity coefficient. 
\subsection{Model B: solutions with the latitudinal angle dependence of $\alpha$}
Upon substituting  $q(\theta)=\tilde{\alpha}$ into the main equations (\ref{1})-(\ref{5}), we obtain 
\begin{equation}
\begin{split}
\frac{v_r^2}{2}+\Omega ^2 \sin ^2\theta =1-\frac{5 c_s^2}{2}-\tilde{\alpha}  (\cot \theta v_r'+\frac{(\rho v_r')'}{\rho}- v_r)
\end{split}
\label{n1}
\end{equation}
\begin{equation}
-\Omega ^2 \sin\theta\cos\theta =-\frac{\left(\rho  c_s^2\right)'}{\rho }+\frac{\tilde{\alpha}  \rho ' v_r}{\rho }+\frac{3\tilde{ \alpha}  v_r'}{2}
\label{n2}
\end{equation}
\begin{equation}
\frac{1}{2} \Omega  v_r=\tilde{\alpha} \Big(\frac{ \rho ' \Omega '}{\rho }+ \Omega ''+3  \cot\theta \Omega '-\frac{3}{4}\Omega\Big)
\label{n3}
\end{equation}
\begin{equation}
-\frac{3 \epsilon  c_s^2 v_r}{2 \tilde{\alpha} }=3 v_r^2+\frac{9}{4} \Omega ^2 \sin ^2\theta +\sin ^2\theta  \left(\Omega '\right)^2+\left(v_r'\right){}^2
\label{n4}
\end{equation}
We are now in a position to solve the above equations using the Fourier analysis method. Before doing so, however, it is insightful to inspect these equations for general trends of the solutions. Equation (\ref{n4}), for instance, shows that its right hand side is always positive for $\gamma<5/3$. It then leads to a negative radial velocity irrespective of the angle $\theta$.

A thorough consideration and classification of the numeric solutions revealed that there are two different branches of solutions: rotating inflow $\Omega\neq 0$, and non-rotating inflow solutions $\Omega=0$. On the other hand each branch can be divided into two main types: solutions which satisfy $v_r(0)=0$, and those which satisfy $v_r(0)=-\epsilon c_s^2/2\tilde{\alpha}$, see equation (\ref{newbc}). Consequently we can categorize our solutions in four different cases:

\subsubsection{Rotating inflow with $v_r(0)=0$}
\begin{figure}
 \center
  \includegraphics[width=93mm]{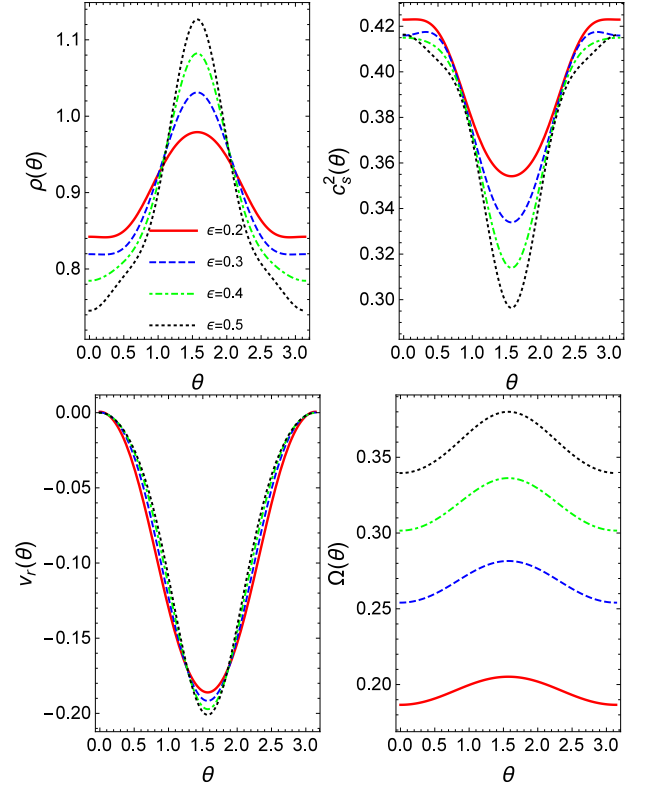}
  \caption{Accretion flow quantities as a function of $\theta$ for $\tilde{\alpha}=0.1$, $\dot{m}=0.23$., $v_{\theta}=0$, $n=3/2$ and different values of the  thermodynamic parameter $\epsilon$.  }
  \label{figure1}
\end{figure}
Fig. \ref{figure1} shows profiles of the physical quantities as a function of $\theta$ with $v_r(0)=0$ for different values of $\epsilon$. The viscosity parameter is $\tilde{\alpha}=0.1$, however, trends of the solutions are the same for other values of $\tilde{\alpha}$. Surface density profile (top, left) shows that its maximum occurs at the equatorial plane. However, the maximum values increases with the thermodynamic parameter $\epsilon$.  These solutions are representative of either ADAFs with $f=1$  different values of $\gamma$ or ADAFs with a constant adiabatic index $\gamma$ and different values of $f$. 

For $\epsilon=0.2$ the density varies by only $13$\% from pole to the equatorial plane. Solutions with small  $\epsilon$, therefore, correspond to nearly spherical flows. For $\epsilon=0.5$, on the other hand, the density contrast between the pole and the equatorial plane is about   $35$\% . Profile of the angular velocity (bottom, right) shows that it is more or less independent of the latitudinal angle, but its values increases with $\epsilon$.  For large values of $\epsilon$, therefore, the system deviates from a spherical symmetry and tends to a rotationally flattened configuration. However, we find that for $\epsilon>1$ the density profile exhibits some oscillations which are not physically plausible. For this reason, we do not report these solutions.
\subsubsection{Rotating inflow with $v_r(0)=-\epsilon c_s^2/2\tilde{\alpha}$}
\begin{figure}
 \center
  \includegraphics[width=93mm]{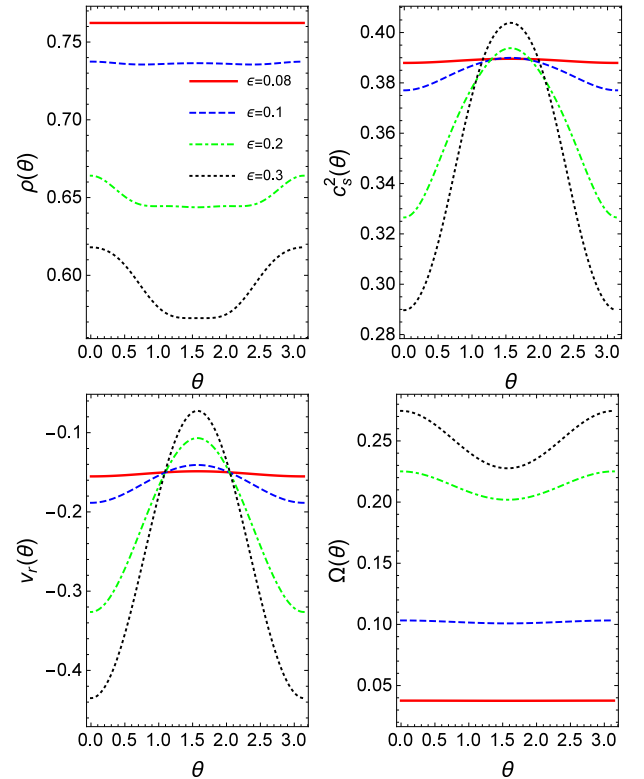}
  \caption{Same as Fig. \ref{figure1}, but for $v_r(0)=-\epsilon c_s^2/2\tilde{\alpha}$.}
  \label{figure2}
\end{figure}
\begin{figure*}
\center
\includegraphics[width=150mm]{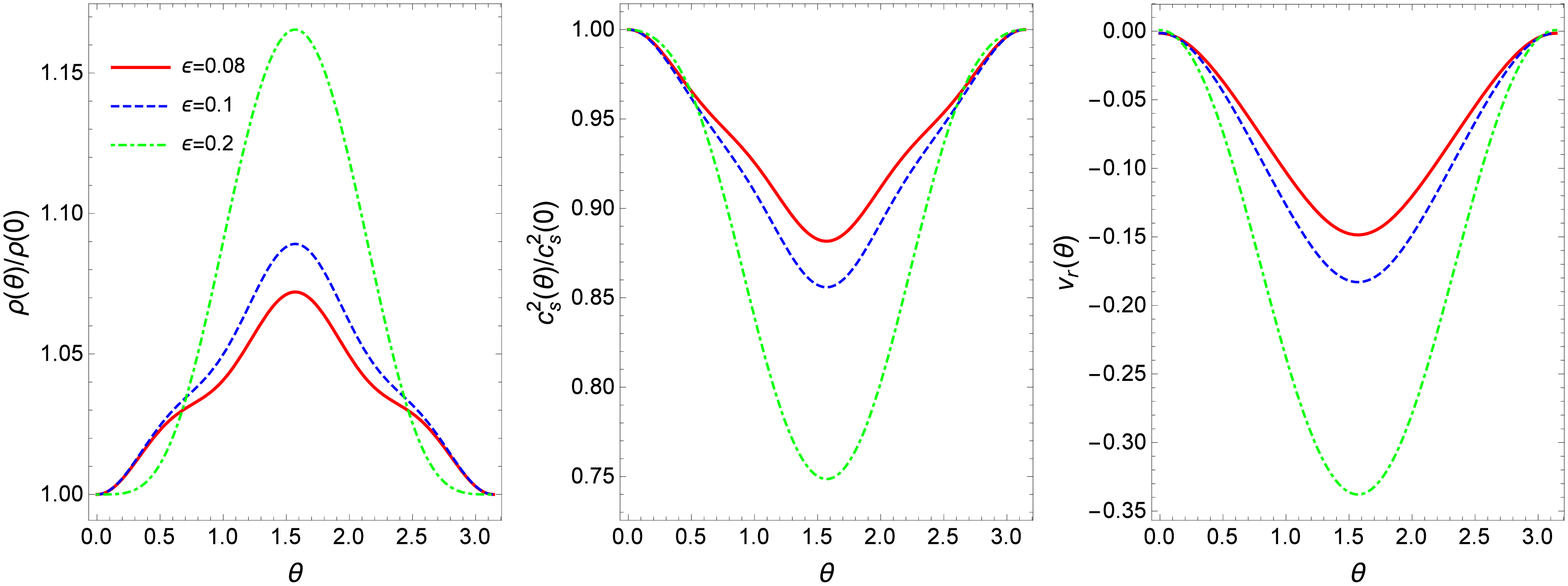}\\
\includegraphics[width=150mm]{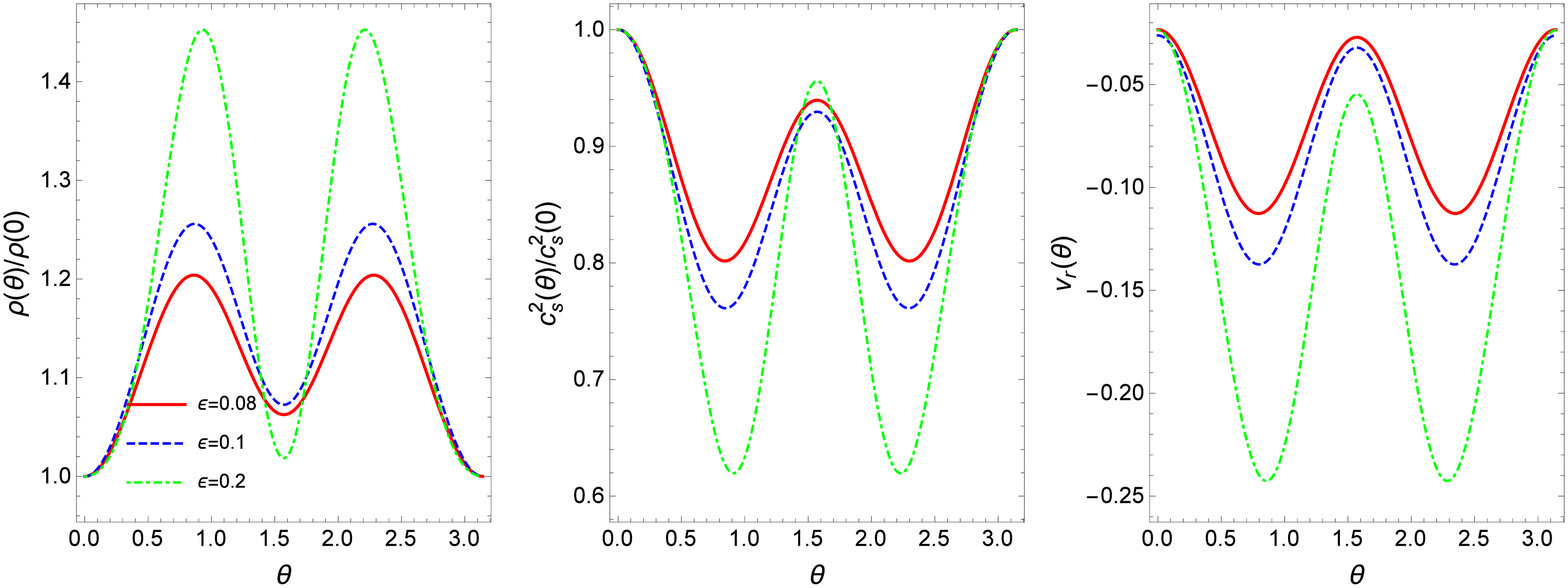}\caption{\textit{Top row:} Non-rotating self-similar solutions for $\tilde{\alpha}=0.1$, $v_{\theta}=0$ and $n=3/2$  and $v_r(0)=0$. \textit{Bottom row:} Non-rotating self-similar solution with the same model parameters as in top row, but with $v_r(0)\neq 0$. Note that here the density and the sound speed are both normalized by their represantitive values at the pole.}\label{figure4}
\end{figure*}
In this particular case, physical quantities are shown in Fig. \ref{figure2}. We note that NY95 found no rotating solution for $v_r(0)\neq 0$, whereas our analysis for the $\alpha$ varying model shows both rotating and non-rotating solutions when the radial velocity does not vanish at the poles. The behavior of the flow in this case is somehow opposite to the case where $v_r(0)=0$. In other words, profiles of the sound speed and the radial velocity reach to a maximum at the equatorial plane, but these quantities  have minimum in Fig. \ref{figure1}. Also, the density and the angular velocity have a  minimum at $\theta=\pi/2$ in Fig. \ref{figure2} while they have maximum at equatorial plane when $v_r(0)=0$.

In both cases, however, the flow deviates from a spherical configuration with increasing the thermodynamic parameter. Nevertheless, we find that this kind of solution exists only for a relatively small range of $\epsilon$, i.e.,  $0.08\lesssim\epsilon\lesssim 0.3$. We note that the range is modified depending upon value of $\tilde{\alpha}$.
\subsubsection{Non-rotating inflow with $v_r(0)=0$}
The top row of Fig. \ref{figure4} shows solutions with $v_r(0)=0$. We find that there is no solution when  $\epsilon\gtrsim 0.5$. Although the inflow does not rotate, its geometrical shape is not purely spherical. This trend is not surprising in the sense that the $\alpha$ viscosity profile is a function of $\theta$. Furthermore, by increasing the thermodynamic parameter $\epsilon$, the system  deviation from a spherical configuration becomes more significant. Density contrast between the pole and the equatorial plane reaches to $17$\% for $\epsilon=0.2$. It should be noted that in the  NY95, there is no non-rotating solution with $v_r(0)=0$. In other words, the non-rotating solutions presented in NY95, possess a negative radial velocity at the poles.
\subsubsection{Non-rotating inflow with $v_r(0)=-\epsilon c_s^2/2\tilde{\alpha}$}
The bottom row of Fig. \ref{figure4} displays solutions with $v_r(0)=-\epsilon c_s^2/2\tilde{\alpha}$. These profiles exhibit different maximum and minimum points which are unlikely to be representative of any physical system. For the sake of completeness, however, we report these solutions. It can be compared with the non-rotating solution presented in NY95, see Appendix B of NY95. Their solution is purely spherical and all functions do not depend on $\theta$. Therefore their solution is reminiscent of the Bondi accretion in the presence of viscosity. However, our solution is not spherical, and similar to other solutions presented in this paper increasing the thermodynamic parameter $\epsilon$, causes more deviation from spherical configuration. It is seen from the left panel in the bottom row of Fig. \ref{figure4} that the density contrast can be more that $45$\%.

However, this deviation from spherical configuration does not mean that there is no spherical non-rotating solution in our varying $\alpha$ model. Let us briefly discuss a simple analytic non-rotating inflow spherical solution. When latitudinal angle dependence of the variables is neglected (i.e., spherical symmetry), we can find non-rotating solution using equations (\ref{n1})-(\ref{n4}). Thus,
\begin{equation}
\begin{split}
&v=\frac{\alpha  (\epsilon +5)-\sqrt{\alpha ^2 (\epsilon +5)^2+2 \epsilon ^2}}{\epsilon }\\&
c_s^2=\frac{2 \alpha }{\epsilon ^2}\left(\sqrt{\alpha ^2 (\epsilon +5)^2+2 \epsilon ^2}-\alpha  (\epsilon +5)\right)
\end{split}
\end{equation}
This solution corresponds to an inflow configuration, irrespective of $\epsilon$. In other words, unlike in NY95, it is not possible to produce wind by choosing negative thermodynamics parameter $\epsilon$.
\subsubsection{Convergence of the solutions}
\begin{figure}
 \center
  \includegraphics[width=85mm]{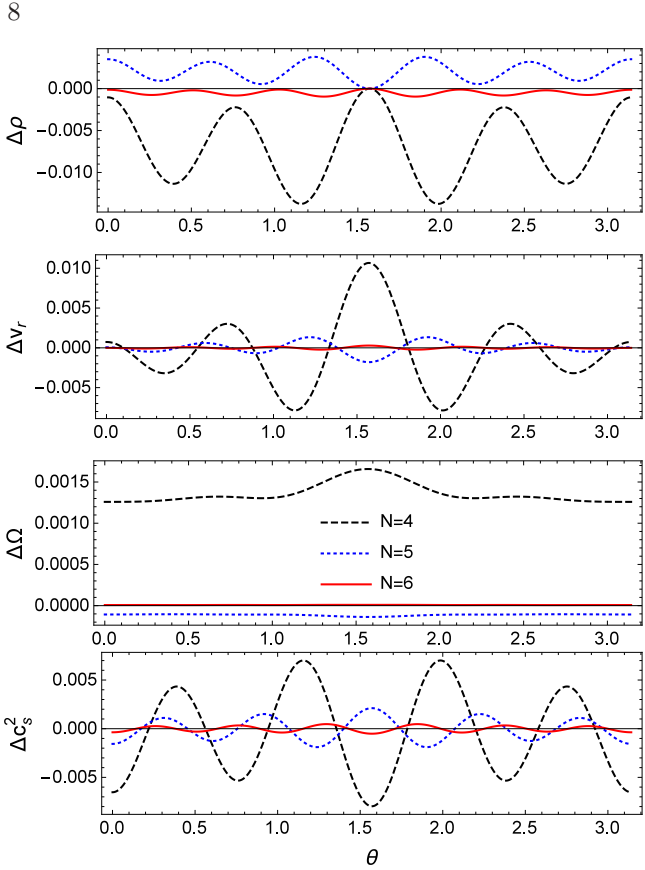}
  \caption{The fractional difference $\Delta Q$ calculated for all physical functions when $\tilde{\alpha}=0.1$ and $\epsilon=0.1$. This figure corresponds to the rotating solutions presented in Fig. \ref{figure1}. Each color belong to a given value of $N$, i.e., number of terms in the Fourier expansion.}
  \label{figure3}
\end{figure}
\begin{figure*}
 \center
\includegraphics[width=150mm]{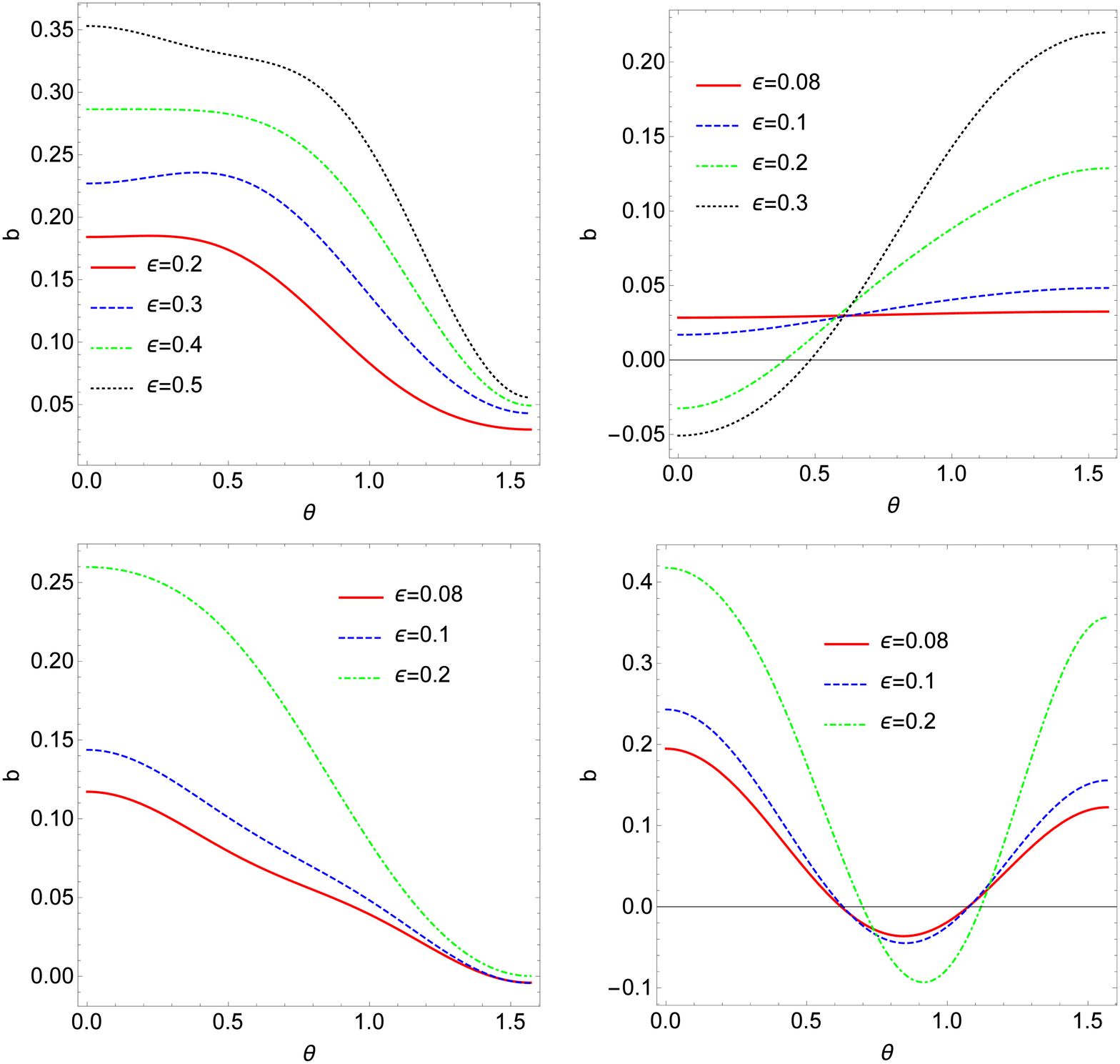}
  \caption{Dimensionless Bernoulli parameter $b$ as a function of $\theta$ for solutions presented in Fig. \ref{figure1}, \ref{figure2} and \ref{figure4}. The top left and right panels correspond to the rotating solutions presented in Fig. \ref{figure1} and \ref{figure2}, respectively. The bottom left and right panels, however, correspond to the non-rotating self-similar solutions presented in the left and the right panels in Fig. \ref{figure4}, respectively.}
\label{figure5}
\end{figure*}
We now verify that the presented solutions are convergent. We increase the number of Fourier terms until the solution converges. In doing so, we  define fractional difference $\Delta Q$ between solutions with $N$ and $N+1$ Fourier coefficients, i.e., 
\begin{equation}
\Delta Q=\frac{Q_{N+1}-Q_N}{Q_{N}}\times 100
\end{equation} 
where $Q$ stand for $\rho$, $v_r$, $c_s^2$ or $\Omega$. Obviously, the solutions are convergent if the fractional difference $\Delta Q$  tends to zero with increasing $N$. Fig. \ref{figure3} shows profiles of $\Delta Q$ for the solutions presented in Fig. \ref{figure1} with $\tilde{\alpha}=0.1$ and $\epsilon=0.1$. Different colors belong to different $N$. It is evident that by increasing $N$, the fractional difference decreases and gets small for $N\geqslant 5$. We generally find that the solutions correspond to fractional differences smaller than $10^{-4}$\% so long as the adopted terms in the Fourier expansions is large than 6.  In other words, the obtained solutions are not modified with increasing the number of terms in the Fourier expansion to larger than 6. It implies  that the solutions are convergent.

\subsection{The Bernoulli parameter:}

In order to study the occurrence of outflow it is important to find Bernoulli function of the flow. This parameter determines the whole energy per unit mass of the flow. As It has been pointed out in NY95 whenever Bernoulli parameter reaches a positive value, one may expect the existence of the outflow in the system. On the other hand, a positive value for Bernoulli parameter means that the flow can escape to infinity as outflow due to its enough energy to overcome gravitational energy. Some researchers have been pointed out the positive values of Bernoulli parameter is a consequence of self-similar solutions \citep{ab,yn99}. Besides, in numerical HD and MHD simulations performed by \cite{yuana,yuanb}, Bernoulli function in the most regions is positive. From this perspective, the Bernoulli parameter $Be$ is useful to check the above mentioned possibility. As usual, let us define the dimensionless parameter $b$ as follows
\begin{equation}
b=\frac{Be}{\Omega_k^2 r^2}=\frac{1}{2}\left(v_r^2+v_{\theta}^2+\sin\theta^2\Omega^2\right)-1+\frac{\gamma}{\gamma-1}c_s^2
\end{equation} 
Fig. \ref{figure5} shows profiles of this parameter for the presented solutions so far. The top left panel in Fig. \ref{figure5} belongs to a rotating solution presented in Fig. \ref{figure1}. We find that within the allowed range of of $\epsilon$, the Bernoulli parameter is always positive. The parameter $b$ becomes larger with  increasing the thermodynamic parameter $\epsilon$. This behavior is completely opposite to the corresponding solution explored in NY95 (i.e., model A) where increasing $\epsilon$ leads to $b<0$ for certain latitudinal angles. This Figure suggests that solutions in Fig. \ref{figure1} are able to produce outflows.

The top right panel in Fig. \ref{figure5} corresponds to the solution presented in Fig. \ref{figure2}. As we have discussed, this solution possesses a non-zero radial velocity at the poles.  Fig. \ref{figure5} shows that for small $\epsilon$, the parameter $b$ is always  positive. For larger values of $\epsilon$, however, the Bernoulli parameter becomes negative in the interval near the poles. We find that this interval gets wider by increasing $\epsilon$. The bottom left panel in Fig. \ref{figure5} shows the Bernoulli parameter for non-rotating solution presented in the top row in Fig. \ref{figure4}. In the case of small $\epsilon$, the parameter $b$ becomes negative at very close to the equatorial plane. 

Finally, the bottom right panel of Fig. \ref{figure5}  belongs to a non-rotating solution displayed in the bottom row of Fig. \ref{figure4}. The Bernoulli parameter is negative within an interval of $\theta$. However it is positive in both near the poles and equatorial plane. There are, therefore, regions where the parameter $b$ is positive and the outflow can exist. However, in order to make sure that the outflow occurs in the system, it is necessary to investigate the convective instability of the solutions. In the next section we study this issue in more details.

\section{Dynamical stability of the solutions}\label{ds}
\begin{figure*}
 \center
\includegraphics[width=150mm]{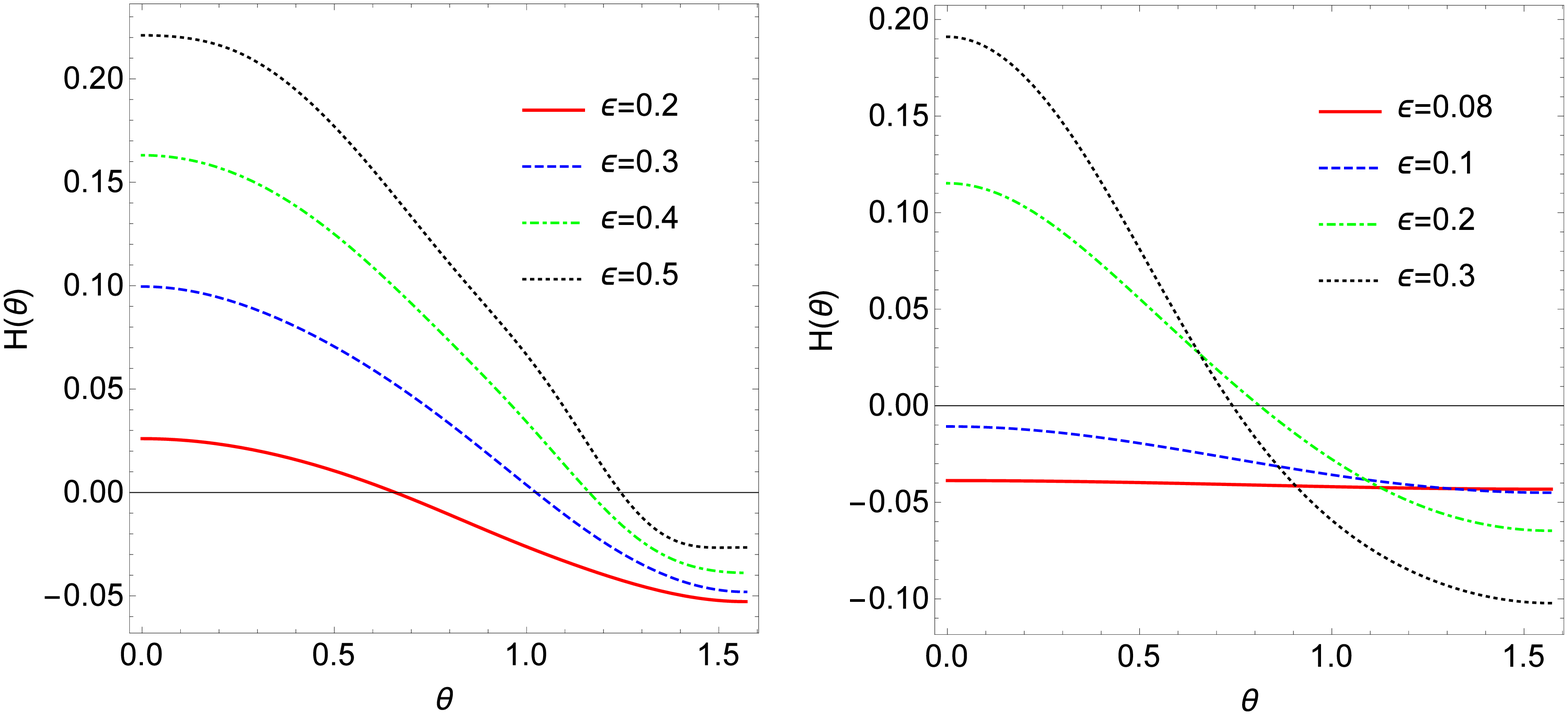}
  \caption{The stability function $H(\theta)$ as a function of $\theta$ for rotating solutions presented in Fig. \ref{figure1},\ref{figure2}. The left and right panels belong to rotating solutions presented in Fig. \ref{figure1} and \ref{figure2} respectively.}
\label{figure6}
\end{figure*}
The stability of the solutions is important in any physical system. For example, it is well-established that the ADAFs with zero $v_{\theta}$ are intrinsically unstable to convective instabilities, see NY95 for more details. As we showed in Appendix B, when $v_{\theta}=0$, there is no outflow in the system. However the convective instability in some regions near the pole can play effectively and the resultant convection outflow can dominate the advection inflow. In other words, this instability can in principle changes the global behavior of the solutions and induce outflow near the pole. From this perspective, and for being sure about the mathematical and physical viability of our solutions, it is necessary to study the response of the system to some pertinent instabilities.

The classic Solberg-H\o iland criteria for stability of the rotating flow against local axisymmetric, adiabatic perturbations in the cylindrical coordinate system $(R,\phi,z)$ are given by \citep{tas}
\begin{equation} \label{hoilad1}
-{1\over \gamma\rho}\del P\bcdot\del\ln P\rho^{-\gamma}
+ {1\over R^3} {\dd R^4\Omega^2\over \dd R} \ge 0,
\end{equation}
\begin{equation}\label{hoilad2}
- {\dd P\over \dd z}\left(
{\dd R^4 \Omega^2\over\dd R} \schwz - {\dd R^4\Omega^2\over\dd
z}\schwR
\right) \ge 0.
\end{equation}

Condition (\ref{hoilad1}) has been widely used to study the convective instability of the astrophysical systems. It is convenient to write it as $N^2+\kappa^2>0$ where $N$ is the usual Brunt-V\"{a}is\"{a}l\"{a} frequency and $\kappa$ is the epicyclic frequency. $N^2$ and $\kappa^2$ are given by the first and second terms on the left- hand side of (\ref{hoilad1}). For a non-rotating flow, the epicyclic frequency is zero and (\ref{hoilad1}) implies the existence of an inward increase of entropy, which is the well-known Schwarzschild criterion. For a rotating flow, the inward increase of entropy is a necessary condition for convective instability. In other words, in a convectively stable flow the entropy decreases inwardly. In an ADAF without radiation, the numerical simulations and analytical descriptions confirm that entropy increases inward. In other words, these systems are convectively unstable, for example see  NY94 and NY95. 

It should be noted that in the regions where the second condition (\ref{hoilad2}) is violated, the local axisymmetric perturbations, in principle, can grow. However, these perturbations are local and can not change the global behavior of the solutions. Our focus, therefore, is on the regions where the first condition (\ref{hoilad1}) is violated. These regions are convectively unstable and can affect the global properties of the flow as explored in NY95 for  the model A.

The left hand side of (\ref{hoilad1}) in the polar spherical coordinate system can be written as $r^{-3/2} H(\theta)$. Therefore the sign of $H(\theta)$ determines the unstable regions. For $H(\theta)<0$,  the latitudinal direction $\theta$ is prone to convective instability. In Fig. \ref{figure6} we display the stability function $H(\theta)$ for the rotating solutions presented in Fig. \ref{figure5}. It should be noted that $H(\theta)$ is negative within the interval $ 0\leq \theta\leq \pi$ for the non-rotating solutions. Therefore the non-rotating solutions are convectively unstable. We can now explore stability of   the rotating self-similar solutions. The left panel of Fig. \ref{figure6} shows that $\epsilon$ has a stabilizing effect on the system in the sense that increasing $\epsilon$ leads to a wider stable interval. However, the stability function $H(\theta)$ is negative near the equatorial plane.

The thermodynamic parameter $\epsilon$ has a stabilizing effect on the rotating solutions with $v_r(0)\neq 0$. The right panel in Fig. \ref{figure6} shows that some parts of the system are stabilized with increasing the parameter $\epsilon$. However, in this case, it does not necessarily extend the stable interval. Similar to the first type of the rotating solutions, we find stability near to the poles and the instability can occur around the equatorial plane.

We showed that all the solutions have regions where $b>0$ and $H(\theta)<0$. This trend, however, does not imply that   convective instability is able to reverse direction of the flow and produce outflows. In other words, as in other hydrodynamic instabilities, the time-scale for the growth of the instability should be small enough compared to other characteristic time-scales in the system. In order to estimate the significance of the convective effects compared to the advection, we follow and generalize the method presented in NY95. To do so,  we assume that $f=1$ and rewrite the energy equation (\ref{n4}) in terms of convective energy flux $\mathbf{F}_c$ and physical functions $c_s^2(r,\theta)$, $v_r(r,\theta)$, $\rho(r,\theta)$ and $\Omega(r,\theta)$ as follows 
\begin{equation}
\begin{split}
&-\frac{3 \epsilon  c_s^2 v_r\rho}{2 r }=-\nabla\cdot \mathbf{F}_c\\& +\frac{\tilde{\alpha}\rho}{\Omega_K}\Big(\frac{3 v_r^2}{r^2}+\frac{9}{4} \Omega ^2 \sin ^2\theta +\sin ^2\theta  \left(\Omega '\right)^2+\frac{\left(v_r'\right)^2}{r^2}\Big)
\end{split}
\label{n40}
\end{equation}
The three terms in this equation, i.e. one term in the left hand side and two terms in the right hand side, are representatives of the advection, the convection and the viscosity respectively (NY95). The system is advection dominated and consequently we need to compare the convection and the advection terms. As in NY95 we assume that the convective flux is proportional to the entropy gradient as
\begin{equation}
\mathbf{F}_c\simeq -K_c \rho(r,\theta) T(r,\theta)\frac{\partial s}{\partial r}\hat{r}
\end{equation}
where $K_c$ is a proportional constant and can be considered as an effective diffusion constant.  Furthermore the specific entropy $s$ and temperature $T$ are given by
\begin{equation}
s=\frac{k_B}{(\gamma-1)m} \ln \frac{p(r,\theta)}{\rho(r,\theta)^{\gamma}}, ~~~T=\frac{m}{k_B}\frac{p(r,\theta)}{\rho(r,\theta)}
\end{equation}
where $k_B$ is the Boltzmann constant and $m$ is the mass of a single molecule, respectively. Now in order to compete this estimation, let us assume that $K_c$ follows a similar profile as the viscosity, i.e. $K_c=\tilde{\alpha}_c/\Omega_K$, where $\tilde{\alpha}_c$  is different from $\tilde{\alpha}$. In fact, in the standard case this parameter can be larger than $\alpha/2$  (NY95). Here, we also assume that $\tilde{\alpha}_c=0.5 \alpha$. Now it is easy to find the ratio of convective term to the advection terms as 
\begin{equation}
G(\theta)=\frac{2 r \nabla\cdot \mathbf{F}}{3 \epsilon  c_s^2 v_r\rho}=-\frac{\tilde{\alpha}_c}{v(\theta)}
\end{equation}
Now the outflow regions can be specified as regions where we have $b>0$,  $H(\theta)<0$ and $\ln G(\theta)>0$. Corresponding to our rotating solutions, however, there is not  a region where all three conditions are satisfied. In other words, although the solutions are convectively unstable, the convective instability can not revert the direction of the flow and cause outflow launching. We note that this is not the case in the model A, and outflow can arise in the rotating solution.

On the other hand, we find  that outflow can happen in the non-rotating solutions. We consider the non-rotating solution with $v_r(0)=0$ (top panel in Fig. \ref{figure4}). In this case, except very close to the equatorial plane, the Bernoulli parameter is positive everywhere. Furthermore, we have  $H(\theta)<0$ for $0\leq\theta\leq\pi$. Therefore we need to check the third condition, i.e. $\ln G(\theta)>0$. This condition is satisfied within the interval $\theta\leq\theta_{\text{crit}}$. The critical angle $\theta_{\text{crit}}$ depends on $\epsilon$. For $\epsilon=0.08$, $0.1$ and $0.2$, the critical angle is found as $\theta_{\text{crit}}=0.6$, $0.53$ and $0.4$ rad, respectively. Our analysis shows that flows with a larger $\epsilon$ are subject to a more collimated outflow.

We also find that outflow can occur in the second type of the non-rotating solutions presented in the bottom panel of Fig. \ref{figure4}. For $\epsilon=0.08$, outflow exists around the poles and the equatorial plane, i.e., within the ranges $\theta<0.29$ and $\theta>1.3$ rad. On the other hand, for $\epsilon=0.1$, the outflow is limited to the regions with  $\theta>1.36$ and $\theta<0.25$. Note that we have written these intervals for $0\leq\theta\leq \pi/2$, and one can simply generalize them to the whole space. For the larger values of $\epsilon$, however, outflow exists only near the poles. When we have $\epsilon=0.2$, for instance, the outflow region is restricted to $\theta<0.2$. It is interesting that although there is no rotation in the system, there is relatively collimated outflow around the poles.

\section{ADAFs with $v_{\theta}(\theta)\neq0$}\label{vnzero}
As we have already mentioned, ADAF solutions with the non-zero $v_{\theta}(\theta)$ have been investigated by XC97 using the same numerical method  implemented by us as well. In this case, the similarity exponent is not necessarily equal to $n=3/2$. We explored a wide parameter space for $n$, $\epsilon$ and $\alpha$, however, our attempts to find solutions with convergent Fourier series were not successful. All obtained solutions which satisfy into the main equations do not fulfill our convergence criterion. We then tried to find convergent solutions by including sine functions in the Fourier series (\ref{fs}). In fact this seems necessary in the sense that without sine functions, the cosine functions can not make a complete orthogonal system. This attempt, however, did not resolve convergence problem.  It is necessary to mention that existence of sine terms does not alter our presented  solutions for $v_{\theta}=0$ because we  found that the sine term coefficients are very small in this case.

The origin of this complexity probably is the intrinsic nonlinear nature of the governing equations when the non-zero $v_{\theta}$ is included. More specifically, when $v_{\theta}$ is zero, the Fourier approach yields to third order algebraic equations. However, for the non-zero $v_{\theta}$, the energy equation leads to fourth order algebraic equations. Also existence of $v_{\theta}$ substantially enlarges the number of terms in each algebraic equations. Consequently, this approach does not work for this case. Therefore we do not confirm the outflow solutions presented in XC97. Furthermore, we think that XC97 solutions have been obtained based on the equations which have some mistakes.
\section{Summary}\label{dis}
In this paper we used the Fourier expansion in order to find semi analytic solutions for ADAFs. More specifically, we assumed that the system is stationary and axisymmetric, and possesses radially self-similar structure. In this case we have a one- dimensional system, and the governing differential equations reduce to an eighth-order system. After setting an appropriate set of boundary conditions, we expand all the physical quantities using the Fourier expansion. In practice one has to truncate the expansions and keep a finite number of terms. In this paper we keep five to nine terms in the expansions. Finally, instead of solving an eighth-order system of differential equations we have solved a set of $5 N$ non-linear algebraic equations to find $5 N$ Fourier coefficients. This means that we find a semi-analytic function for all the physical quantities. The main practical benefit of this approach compared to numerical integration of the differential equations, is that one obtains all the functions in the whole space. This makes easy to study the properties of the system and straightforwardly interpret the results. For example the stability issues of the flow can be easily checked. We remind that one of the restrictions of the numerical integration of the governing equations is that one can not start from the equatorial plane and reach the pole. In brief one may say that the Fourier expansion analysis leads to analytical solutions, and analytical solutions are always helpful to simplify the analysis of the given system. 

Using this approach for a new viscosity model, in which $\alpha$ varies with $\theta$, we have found four categories of the solutions in the absence of the  latitudinal component of the velocity. The first rotating solution presented in Fig. \ref{figure1} corresponds to an inflow  with $\tilde{\alpha}=0.1$ and $v_r(0)=0$. Although this solution is convectively unstable, the convection can not reverse the direction of the flow. The second rotating and inflow solution is illustrated in Fig. \ref{figure2} with $\tilde{\alpha}=0.1$ and $v_r(0)\neq 0$. Convection can not reverse the direction of the flow In this solution too.

The third solution  presented in the top row of Fig. \ref{figure4} is a non-rotating inflow with $\tilde{\alpha}=0.1$ and $v_r(0)= 0$. Although the flow does not  rotate, its geometrical shape is not spherical and it tends to a flattened configuration with increasing $\epsilon$. Furthermore, we showed that convection is dynamically important  and it may contribute to launching of the  outflows. More importantly outflow can exist around the poles. Our last solution, which corresponds to a non-rotating inflow with $\tilde{\alpha}=0.1$ and $v_r(0)\neq 0$, has been shown in the bottom row of Fig. \ref{figure4}. We showed that convection in the system can produce outflows. For small $\epsilon$ outflow exists near the poles and also near the equatorial plane. For large values of $\epsilon$, however, outflow happens only around the poles.

Finally, we studied a case with $v_{\theta}\neq 0$. In this case we could not find any convergent and unique solutions due to highly non-linear nature of the equations. We had attempts  to generalize the method in various directions, however,   we could not find convergent solutions. Consequently we do not confirm the outflow solutions already reported in XC97.

As the final remark, we would like to mention that in this paper we showed that the Fourier expansion method can help to study ADAF systems. Naturally more careful and physically oriented investigations can be accomplished by taking into account more physics in the system. For example one may add magnetic fields, the effect of thermal conduction or existence of the radiation cooling and use this method to derive the properties of the system. It is also possible, in practice, to use this expansion even in the radial direction. In this case one may use radial eigen functions of the Laplace operator. In other words one may study solutions which are not necessarily self-similar in the radial direction. It is even possible to study self-gravitating systems in which the central mass potential deviates from the standard Newtonian potential. For example one may use the sudo-Newtonian potential in order to include the relativistic effects. Therefore more investigation is required to check the effectiveness of this approach.

Given these facts, the treatments in the paper are sufficiently general to describe many disk-wind substructures such as inflow-outflow regions, corona, disk jet and collimated jet which have been appeared in simulations and generally supposed to play important role in power spectrum of the system. The numerical approach presented by NY95 for vertical structure of disk is unable take into account these sub-structures because of complexity of numerical techniques. But using Fourier analysis we will be able to investigate vertical structure uniquely by adding proper physics.

\acknowledgments
We would like to appreciate the anonymous referee who helped us to improve this paper very substantially. Also AH and SA would like to thank Mahmood Roshan for useful comments and discussions. This work was supported by the Ferdowsi University of Mashhad under grant no. 3/44301 (1396/05/24).
\section*{Appendix A:}
In this appendix we show the details of the Fourier approach for a toy model. In fact we set $\Omega=0$ in the main equations (\ref{n1})-(\ref{n4}) and for simplicity we keep only two terms in the expansions. In this case equation (\ref{n3}) is automatically satisfied and we deal with three differential equations (\ref{n1}), (\ref{n2}) and (\ref{n4}). Furthermore the physical variables are given by 
\begin{equation}
\begin{split}
&\rho(\theta)= a_0 + a_1 \cos 2\theta\\&
c_s^2(\theta)=w_0+w_1 \cos 2\theta \\&
v(\theta)= b_0+b_1 \cos 2\theta
\end{split}
\label{a4}
\end{equation} 
Substituting these functions into equations (\ref{n1}), (\ref{n2}) and (\ref{n4}) we find the following equations respectively
\begin{equation}
\begin{split}
& 2 a_1 \Big(2 b_0 \cos 2 \theta  \Big(b_1 \cos 2 \theta-\alpha \Big)-\alpha  b_1 (4 \cos 2 \theta\\&+11 \cos 4 \theta +3)+b_1^2 \cos ^3 2 \theta +b_0^2 \cos 2 \theta +\cos 2 \theta \times\\& \Big(5 w_1 \cos 2 \theta +5 w_0-2\Big)\Big)+a_0 \Big(-4 b_0 \Big(\alpha -b_1 \cos 2 \theta \Big)\\&-4 \alpha  b_1 (7 \cos 2 \theta +2)+b_1^2 (\cos 4 \theta +1)+2 b_0^2\\&+2 \Big(5 w_1 \cos 2 \theta +5 w_0-2\Big)\Big)=0
\end{split}
\label{nn1}
\end{equation}
\begin{equation}
\begin{split}
&\sin \theta \cos ^3 \theta  \Big(60 \alpha  a_1 b_1-48 a_1 w_1\Big)+\cos \theta  \Big(\sin ^3 \theta \times\\& \Big(48 a_1 w_1-60 \alpha  a_1 b_1\Big)+\sin \theta \times \\& \Big(24 \alpha  a_1 b_0+36 \alpha  a_0 b_1-24 a_1 w_0-24 a_0 w_1\Big)\Big)=0
\end{split}
\label{nn2}
\end{equation}
\begin{equation}
\begin{split}
&12 \alpha  b_0^2+14 \alpha  b_1^2+\sin ^4 \theta  \Big(3 b_1 w_1 \epsilon -2 \alpha  b_1^2\Big)+\sin ^2 \theta\times\\&  \Big(-24 \alpha  b_0 b_1-6 b_1 w_0 \epsilon -6 b_0 w_1 \epsilon \Big)+\cos ^4 \theta  \Big(3 b_1 w_1 \epsilon\\& -2 \alpha  b_1^2\Big)+\cos ^2 \theta  \Big(24 \alpha  b_0 b_1+\sin ^2 \theta  \Big(12 \alpha  b_1^2-18 b_1 w_1 \epsilon \Big)\\&+6 b_1 w_0 \epsilon +6 b_0 w_1 \epsilon \Big)+6 b_0 w_0 \epsilon +3 b_1 w_1 \epsilon=0
\end{split}
\label{nn3}
\end{equation}
Now we rewrites products and powers of sine and cosine functions in terms of trigonometric functions with combined arguments. Therefore equations (\ref{nn1})-(\ref{nn3}) take the following form
\begin{equation}
\begin{split}
&A_1+A_2\cos 2 \theta +A_3\cos 4 \theta +a_1 b_1^2 \cos 6 \theta=0\\&
B_1 \sin 2 \theta +B_2 \sin 4 \theta=0\\&
C_1+C_2 \cos 2 \theta  +C_3 \cos 4 \theta=0 
\end{split}
\end{equation}
where the coefficients are defined as
\begin{equation}
\begin{split}
&A_1=-8 \alpha  a_0 b_0-16 \alpha  a_0 b_1-12 \alpha  a_1 b_1 +4 a_0 b_0^2+\\&~~~~~~+4 a_1 b_1 b_0+2 a_0 b_1^2+20 a_0 w_0+10 a_1 w_1-8 a_0\\&
A_2=-8 \alpha  a_1 b_0-56 \alpha  a_0 b_1-16 \alpha  a_1 b_1+4 a_1 b_0^2\\&~~~~~~+8 a_0 b_1 b_0+3 a_1 b_1^2+20 a_1 w_0+20 a_0 w_1-8 a_1\\&
A_3=-44 \alpha  a_1 b_1+2 a_0 b_1^2+4 a_1 b_0 b_1+10 a_1 w_1\\&
B_1=4 \alpha  a_1 b_0+6 \alpha  a_0 b_1-4 a_1 w_0-4 a_0 w_1\\&
B_2=  5 \alpha  a_1 b_1-4 a_1 w_1\\&
C_1=12 \alpha  b_0^2+14 \alpha  b_1^2+6 b_0 w_0 \epsilon +3 b_1 w_1 \epsilon\\&
C_2=24 \alpha  b_0 b_1+6 b_1 w_0 \epsilon +6 b_0 w_1 \epsilon\\&
C_3=3 b_1 w_1 \epsilon -2 \alpha  b_1^2
\end{split}
\end{equation}
Considering that we have kept only two terms in the expansions, now we need to set to zero the coefficients of $\cos m\theta$ and $\sin m\theta$ for $m=0$ and $m=2$. Therefor for a given $\alpha$ and $\epsilon$, we have the following five algebraic equations for six unknowns $a_0$, $a_1$, $b_0$, $b_1$, $w_0$ and $w_1$: 
\begin{equation}
A_1=0, ~~A_2=0,~~B_1=0,~~C_1=0,~~C_2=0
\label{a1}
\end{equation}
On the other hand, as discussed in the subsection \ref{bcond}, we have one more equation from the boundary condition $\dot{m}=0.23$. This constraint is given by 
\begin{equation}
-2 a_0 b_0+\frac{2 a_1 b_0}{3}+\frac{2 a_0 b_1}{3}-\frac{14 a_1 b_1}{15}=0.23
\label{a2}
\end{equation}
Finally, equations (\ref{a1}) and (\ref{a2}) are six algebraic equations for six unknowns, and can be solved using usual numeric procedure. Then we choose the physical solutions among the solutions. Consequently, although the coefficient are obtained using numerical methods, we have semi-analytic solutions given by (\ref{a4}). For larger number of terms in the expansions, we use the same procedure to solve the main differential equations. 
\begin{figure}
\center
\includegraphics[width=87mm]{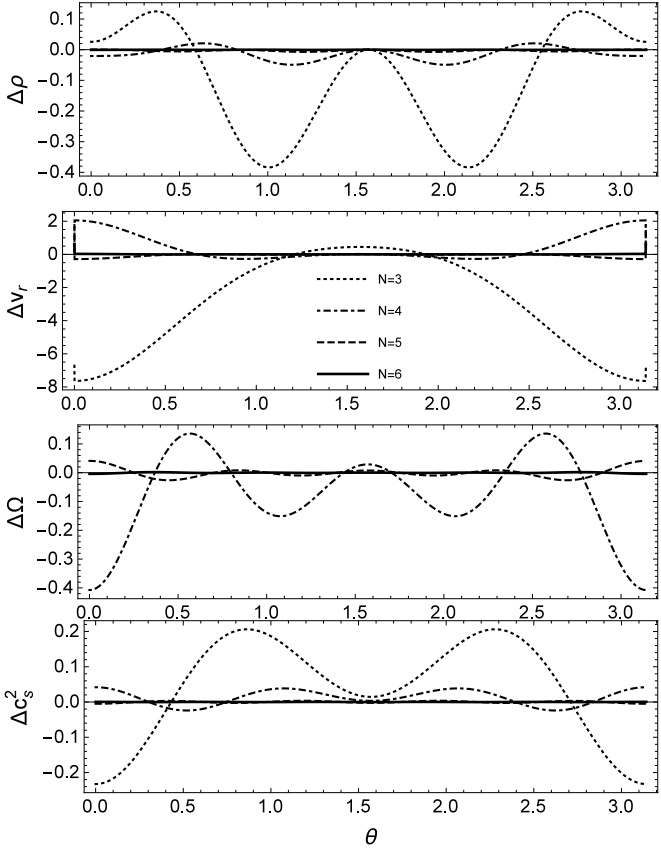}
\caption{Convergence of the solutions by increasing the number of terms in the Fourier expansion. As it is clear the fractional difference between solutions obtained with keeping $N$ and $N+1$ terms in the expansions, gets very small by increasing $N$.}
\label{conver2}
\end{figure}

\begin{figure}
\includegraphics[width=85mm]{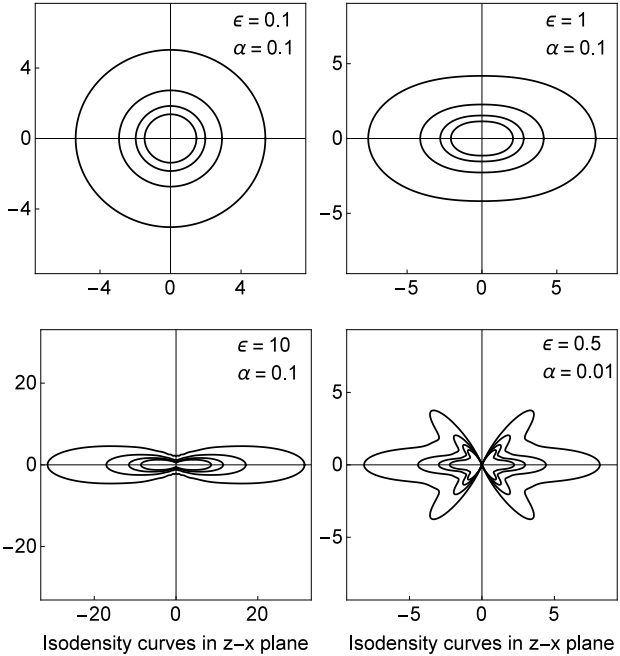}
\caption{Isodensity contours in the $z-x$ plane for four solutions. The top panels and the bottom left panel belongs to the self-similar solutions illustrated in Fig. \ref{fig1}. The bottom right panel belongs to the small $\alpha$ case which correspond to a higher order solution. }
\label{fig2}
\end{figure}
\begin{figure*}
 \center
  \includegraphics[width=130mm]{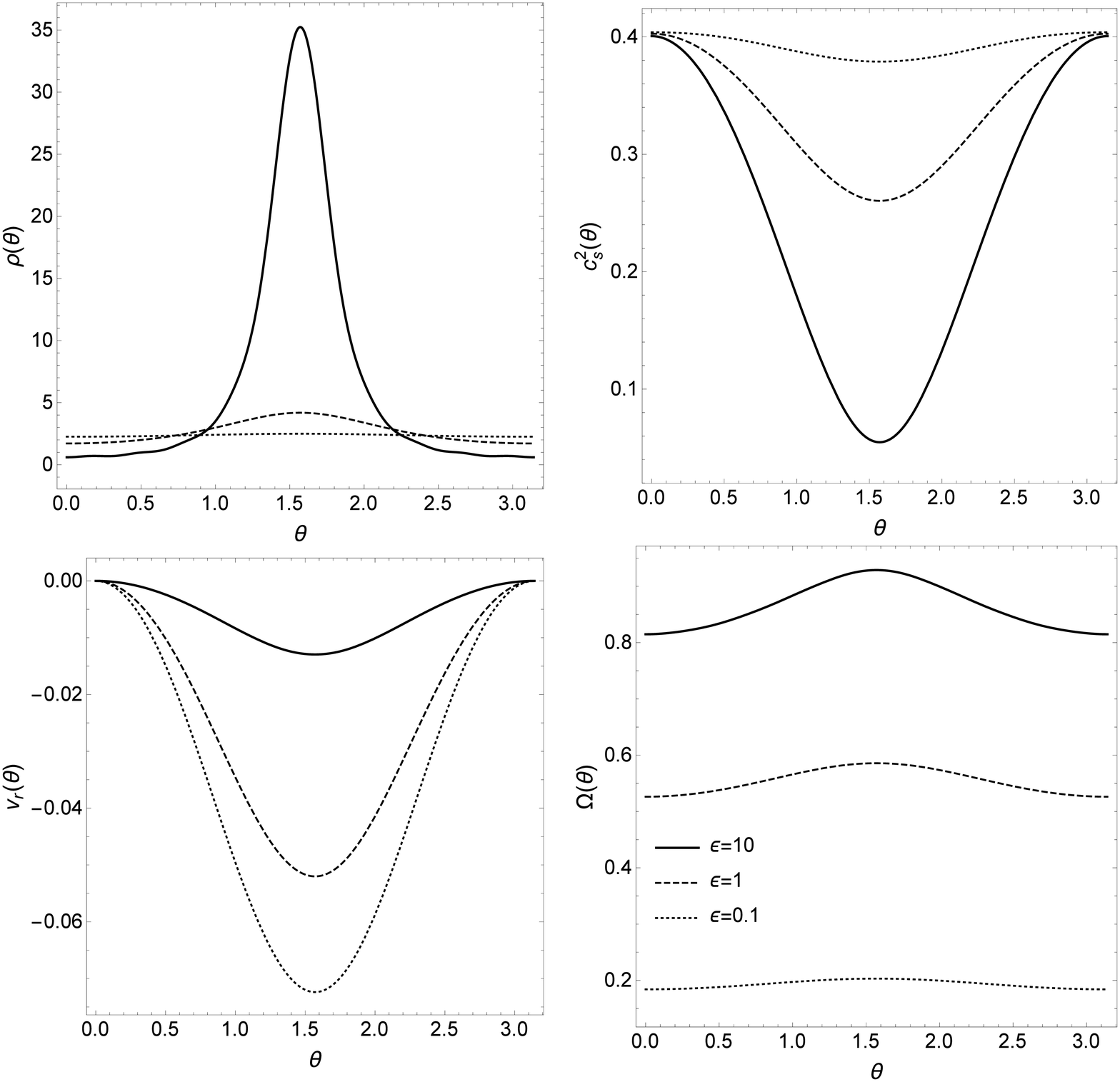}
  \caption{
  Self-similar solutions corresponding to $n=3/2$, $\alpha=0.1$, $\epsilon=0.1, 1, 10$ and $\dot{m}=0.23$. \textit{Top left}: density $\rho(\theta)$. \textit{Top right}: the isothermal sound speed $c_s^2$ with respect to $\theta$. \textit{Bottom left}: radial velocity $v_r(\theta)$. \textit{Bottom right}: the angular velocity $\Omega(\theta)$.}
 \label{fig1}
\end{figure*}
\section*{Appendix B: Retrieving NY95 results} \label{NY95}

In order to check the validity and correctness of Fourier analysis of ADAFs, we try to recover NY95 results. We set $n=3/2$ and consequently $v_{\theta}(r,\theta)=0$. In this case, it is straightforward to show that the differential equations are sixth order and we need six boundary conditions. As we mentioned, in this case, we use the net mass accretion rate to obtain one boundary condition as $\dot{m}=0.23$, as what has been done in NY95. Albeit magnitude of $\dot{m}$ has not explicitly reported in NY95. Here we chose $\dot{m}=0.23$ in order to find a lost relation to those of NY95. The boundary conditions (\ref{bc1}) and (\ref{bc2}) are automatically satisfied. We investigate both conditions on $v_r(0)$ and compare the results with those presented in NY95 which have been obtained by a numerical relaxation technique. It is necessary to mention that there are two free parameters which control the physics of the system: $\alpha$ and $\epsilon$. In the following, we briefly report our results, and the physical interpretations of the solutions can be found in the comprehensive paper NY95.

Let us start with $v_{r}(0)=0$. It turns out that for $N\geq 6$ solutions are convergent. We have shown this fact in Fig. \ref{conver2}. More specifically, to see the convergence of the solutions, we have plotted the fractional difference between solutions obtained with retaining $N$ and $N+1$ terms in the expansions. It is clear that by increasing the number of terms $N$ the fractional differences in all the physical quantities get small. In fact for $N\geq 6$ the fractional differences are almost zero. 

We present the results for $N=9$. In other words, we keep ten terms in the Fourier expansions. As mentioned in NY95, it turns out that this boundary condition belongs to rotating, i.e. $\Omega\neq 0$, and fully advective solutions.  The isodensity contours in the meridional plane have been illustrated in Fig. \ref{fig2}. This figure can be compared with Fig. 2 in NY95. For $\alpha=0.1$ we have plotted the quantities for different values of $\epsilon$ in Fig. \ref{fig1}. This figure should be compared with the Fig. 1 in NY95.

\begin{figure}
\center
\includegraphics[width=85mm]{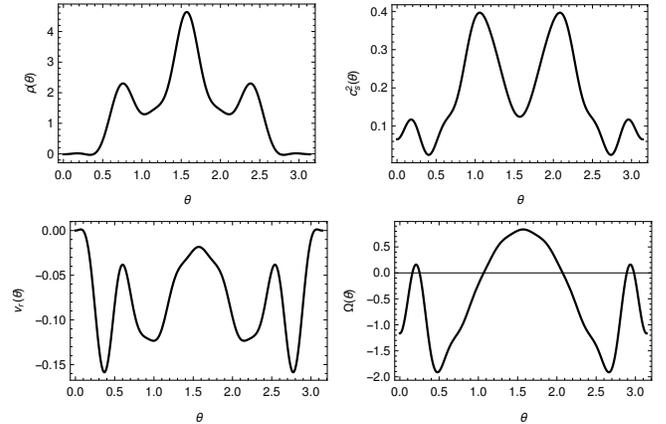}
\caption{Physical quantities for a higher-order solution for which $n=3/2$, $\alpha=0.01$, $\epsilon=0.5$ and $\dot{m}=0.23$. \textit{Top left}: density $\rho(\theta)$. \textit{Top right}: the isothermal sound speed $c_s^2$ with respect to $\theta$. \textit{Bottom left}: radial velocity $v_r(\theta)$. \textit{Bottom right}: the angular velocity $\Omega(\theta)$.}
\label{fig3}
\end{figure}

As reported in NY95 for small $\alpha$ there are some higher-order solutions in them the angular velocity reverses sign one or more times as a function of $\theta$. These solutions are unlikely to describe a real system. In NY95 only the isodensity contour of such a solution has been reported. For completeness, we have also found a higher-order solution for $\alpha=0.01$ and $\epsilon=0.5$ and shown the relevant quantities in Fig. \ref{fig3}. The corresponding isodensity contour has been shown in the bottom right panel of Fig. \ref{fig2}. As it is clear from Fig. \ref{fig3}, $\Omega$ changes sign six times in the interval $[0,\pi]$ while $v_r$ is oscillatory but always negative.

So far it is clear that there is an excellent agreement between our results and those presented in NY95. Finally, the only class of solutions which we have to compare with NY95 is the solutions for which $v_r(0)=-\epsilon/2\alpha$. Applying the Fourier series approach to this case, we found that there is no rotating solution. For example for $\alpha=0.1$, $\epsilon=0.1$ and $\dot{m}=0.23$ we found that $\rho$, $v_r$ and $c_s^2$ are constant and are given by $0.23$, $-0.5$ and $0.34$ respectively. On the other hand, $\Omega(\theta)$ oscillates in the narrow interval $-2.4\times 10^{-8}<\Omega<3.2 \times 10^{-8}$, which is negligible compared to other velocity components. Our conclusion for non-existence of rotation solutions for this boundary condition is completely in agreement with the analytical description presented in Appendix of NY95.

\begin{figure}
\center
\includegraphics[width=85mm]{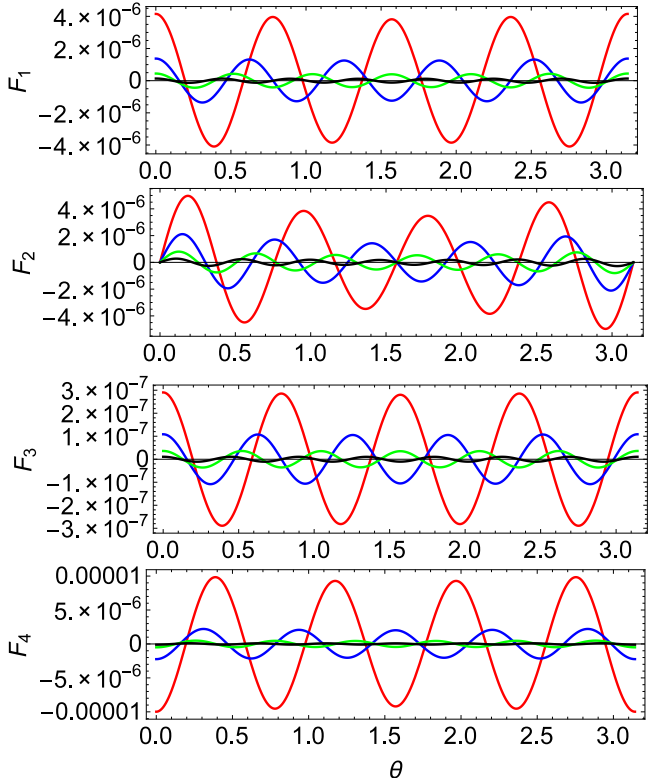}
\caption{Functions $F_i$ in terms of $\theta$ for different values of $N$ when $\epsilon=\alpha=0.1$ in our second rotating solution. The red, blue, green and black curves belong to $N=4$, $N=5$, $N=6$ and $N=7$ respectively.}
\label{figure7}
\end{figure}

With this test, we checked the reliability of the Fourier series approach. However let us introduce another more direct test to show that this method leads to true solutions for the main equations. To do so we rewrite the equations (\ref{n1})-(\ref{n4}) as follows respectively
\begin{equation}
F_1(\theta)=0,~~ F_2(\theta)=0,~~ F_3(\theta)=0,~~ F_4(\theta)=0
\label{a5}
\end{equation}
One may note that this method leads to semi analytic solutions. For example for our second type of solutions with $\alpha=0.1$ and $\epsilon=0.1$, when $N=4$ the function $\rho(\theta)$ is given by
\begin{equation}
\rho(\theta)\simeq1+(4.7 \cos 2 \theta +9.2 \cos 4 \theta +2.1 \cos 6 \theta)\times 10^{-4}\nonumber
\end{equation}
we have similar expansions for other physical variables. Therefore if we substitute them into the main equations (\ref{a5}), we expect that the right hand side of all equations in (\ref{a5}) vanishes. For different values of $N$ and for the above mentioned solution, we have plotted functions $F_i(\theta)$ in Fig. \ref{figure7}. As it is clear from this figure, these functions get smaller by increasing $N$. When $N=7$ all functions are smaller than $10^{-7}$ in the whole space, i.e. $|F_i(\theta)|<10^{-7}$ in $0<\theta\leq \pi$. These functions get even smaller for larger choices of $N$. This behavior is consistent with the convergence of the solutions presented in Fig. \ref{figure3}. Therefore one can conclude that solutions obtained with this Fourier approach, are true solutions for the main differential equations.

\bibliographystyle{apj}
\bibliography{short,FADAF2}
\clearpage
\end{document}